\documentclass[prl,twocolumn,superscriptaddress]{revtex4}
\usepackage{epsfig}
\usepackage{times}
\usepackage{amsmath}
\usepackage{amsfonts}
\usepackage{amssymb}
\usepackage[usenames]{color}
\usepackage{graphicx}

\newcommand{\beq}{\begin{equation}}
\newcommand{\eeq}{\end{equation}}
\newcommand{\beqa}{\begin{eqnarray}}
\newcommand{\eeqa}{\end{eqnarray}}

\begin{document}

\title{Probabilistic growth of large entangled states with low error accumulation}
\author{Yuichiro Matsuzaki}
\affiliation{Department of Materials, University of Oxford, OX1 3PH, U. K.}
\author{Simon C. Benjamin}
\affiliation{Department of Materials, University of Oxford, OX1 3PH, U. K.}
\affiliation{Centre for Quantum Technologies, National University of Singapore, 3 Science Drive 2, Singapore 117543.}
\author{Joseph Fitzsimons}
\affiliation{Department of Materials, University of Oxford, OX1 3PH, U. K.}
\affiliation{Institute for Quantum Computing, University of Waterloo, Waterloo, Ontario, Canada}

\begin{abstract}
The creation of complex entangled states, resources that enable quantum computation, can be achieved via simple `probabilistic' operations which are individually likely to fail. However, typical proposals exploiting this idea carry a severe overhead in terms of the accumulation of errors. Here we describe an method that can rapidly generate large entangled states with an error accumulation that depends only logarithmically on the failure probability. We find that the approach may be practical for success rates in the sub-$10\%$ range, while ultimately becoming unfeasible at lower rates. The assumptions that we make, including parallelism and high connectivity, are appropriate for real systems including measurement-induced entanglement. This result therefore shows the feasibility for real devices based on such an approach.
\end{abstract}

\maketitle

The problem of scalability continues to be a key challenge in the field of quantum information processing (QIP). Whereas many physical systems have successfully embodied a few qubits, a clear route toward large scale universal computers has yet to be demonstrated. One promising solution is {\em distributed} QIP, where small systems (such as trapped atoms or solid state nanostructures~\cite{Benjamin:2009p374}) are networked together to constitute the entire machine. While this may resolve the issue of scaling, it introduces the problem of how to entangle the physically remote subsystems. Solutions were found~\cite{Cabrillo:1999p339,Bose:1999p326} involving the use of optical measurements that simultaneously monitor two, or more~\cite{Benjamin:2005p362}, such systems. There are now experimental demonstrations of such approaches in both ensemble systems~\cite{Chou:2005p333} and with individual atoms~\cite{Moehring:2007p337}.

Typically a remote entangling operation (EO) has two key characteristics: First, it may fail outright, but this failure will be {\em heralded}, meaning that the failure will be registered by the apparatus. Failure is destructive, leaving the qubits that were acted upon in an uncertain state so that they will need to be reset. We should assume that such failures are common, i.e. that the success probability $p_s$ may be very low in real systems. For example, in the work of Monroe {\em et al.} impressive proof-of-principle experiments have achieved entanglement by measurement of two remote atoms, but the success rate is below one in a million~\cite{Moehring:2007p337, MonrePRL08}. The second characteristic of a realistic EO will be some finite probability of {\em un-herladed} errors, including all imperfections in the operation that are unrecorded by the apparatus. 

\begin{figure}[!]
\includegraphics[width = \columnwidth]{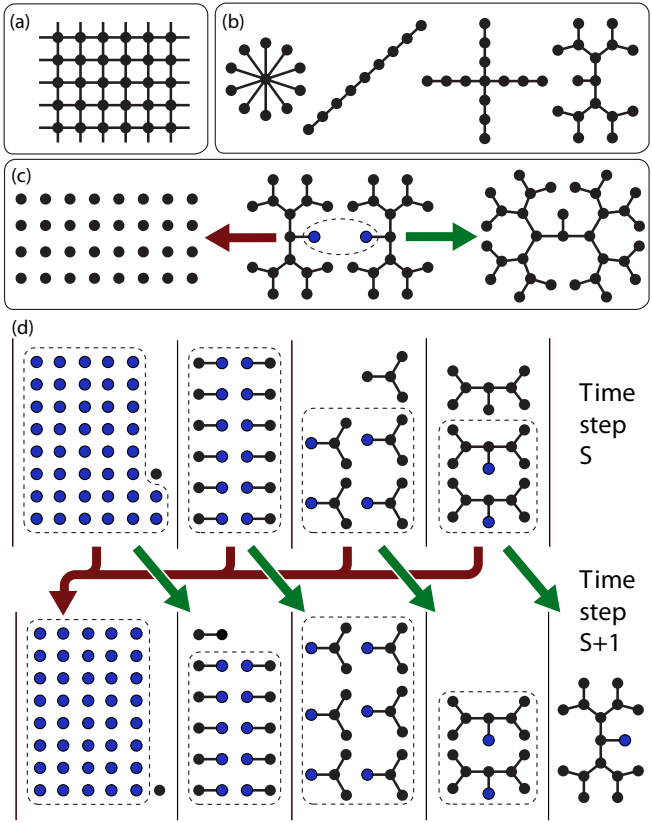}
\caption{The figure shows {\em graph states}: nodes correspond to qubits, and connections (`edges') correspond to phase entanglement. (a) The cluster state is a resource to permit general quantum information processing. (b) Several kinds of graph state have been considered as `building blocks' including star~\cite{Nielsen:2004p371}, linear~\cite{Barrett:2005p363} and cross~\cite{Duan:2005p369} geometries. However, the `snowflake' that we introduce (rightmost) offers superior error suppression. (c) When we attempt to entangle two qubits (blue) from two independent snowflakes, we either succeed to form a single new snowflake (green arrow), or we fail and completely reset all qubits (red arrow). (d) An example of the snowflakes that will exist within our device in two {\em consecutive} time steps. At each step we pair up snowflakes of equal size, and attempt to fuse all such pairs in parallel (specific qubits to be involved in entanglement operations are marked blue).   }
\label{fig:simpleFusion}
\end{figure}

The challenge is then to create a large scale entangled state across the network using a basic EO of this kind. To be specific one might aim to create a so-called {\em graph state} (Fig.~\ref{fig:simpleFusion}). These states are conveniently represented diagrammatically, with nodes corresponding to qubits and lines (or `edges') representing phase entanglement between them. Graph states with certain topologies can enable quantum computing because they incorporate all the entanglement required to perform an algorithm; Fig.~\ref{fig:simpleFusion}(a) shows one suitable example, a square lattice called a {\em cluster state}. A cluster state is therefore an example of the kind of large entangled state one might wish to create; we refer to such targets generically as `the primary graph state'.


Creating a network-wide graph state though the use of failure-prone EOs is actually quite straightforward if each local subsystem of the network contains two or more qubits; then we can use one complete subset to store the growing graph state, while the other set is involved in `brokering' new entanglement~\cite{BriegelDur03,Benjamin:2006p358}. Unfortunately, many physical systems may be limited to embodying only a single qubit. Given only one qubit at each network site, it is inevitable that the nacent graph state will be damaged repeatedly during its creation: every time we wish to entangle two specific qubits, there is a significant risk that the EO will fail and therefore the two qubits in question will need to be reset, losing any prior entanglement they had acquired with third party qubits.  At first glance it may seem that it will not be possible to grow a large state efficiently unless the probability of success is at least $0.5$ (say). However, previous publications have shown that one can indeed achieve positive growth {\em on average} for any finite $p_S$~\cite{Lim:2005p364,Barrett:2005p363,Nielsen:2004p371,Duan:2005p369,Rohde:2007p370}. Generally the solution involves small resource states (see Fig.~\ref{fig:simpleFusion}(b)) which are created by `brute force', i.e. suffering the cost of repeated failures. Then these small graph states are added to the primary graph state. When a small state is successfully connected to the primary graph, the several qubits that are thus added {\em more than make up for} those which are lost though failures. 

This kind of strategy has two drawbacks, both of which we address in the present manuscript. Firstly, the time cost of creating the smaller `resource' object will be very high when $p_S$ is small. This cost cannot be ignored, since (for a finite number of physical qubits) it dictates the maximum rate at which the primary graph can be grown; if this rate is too low, decoherence will destroy the primary graph before it can be completed. While one cannot avoid this time cost, we believe that the approach that we report here is the most efficient to have been described to-date. The second drawback of the use of resource objects, often overlooked in earlier works, concerns the accumulation of errors. Such errors are broadly of two kinds: The complexity of the resource states is in itself a source of errors, since imperfections in the `successful' entanglement operations will accumulate in the resource objects, and ultimately lead to degradation of the primary graph state. Furthermore, the large time cost associated with the resource preparation may imply that passive decoherence (as opposed to errors from active operations) will significantly degrade the resource during its formation, and again the primary graph would inherit this noise. In the protocol we describe here, both forms of errors accumulate only as a logarithmic function of $1/p_S$. We believe that our protocol is the first to have a logarithmic error accumulation, whereas in previous approaches errors accumulate linearly. In this sense the present scheme is considerably more practical.

As with previous proposals, our protocol is based on the creation of relatively small resource graph states, which are fused together to create the primary graph state. Earlier schemes have used linear, star or cross-shaped topologies for the resource objects (Fig~\ref{fig:simpleFusion}). All these topologies have the property of redundancy: the structure contains a number of qubits of order $1/p_S$ so that the effort to join the resource into the primary graph state can suffer multiple failures prior to the success. Regrettably the primary graph will ultimately accumulate errors corresponding to this redundant structure, either during the process of attaching the resource graph state or in subsequent `pruning' of the remaining redundant structure after success. Here we employ a different topology; it is simply a balanced binary tree but we refer to it as a {\em snowflake} since we find it helpful to envisage it as roughly circular (see Fig.~\ref{fig:simpleFusion}). This structure is chosen because (a) it is efficient to grow, and moreover (b) only a logarithmic fraction of imperfections during growth eventually afflict the primary graph state. 

\begin{figure}[!]
\includegraphics[width = \columnwidth]{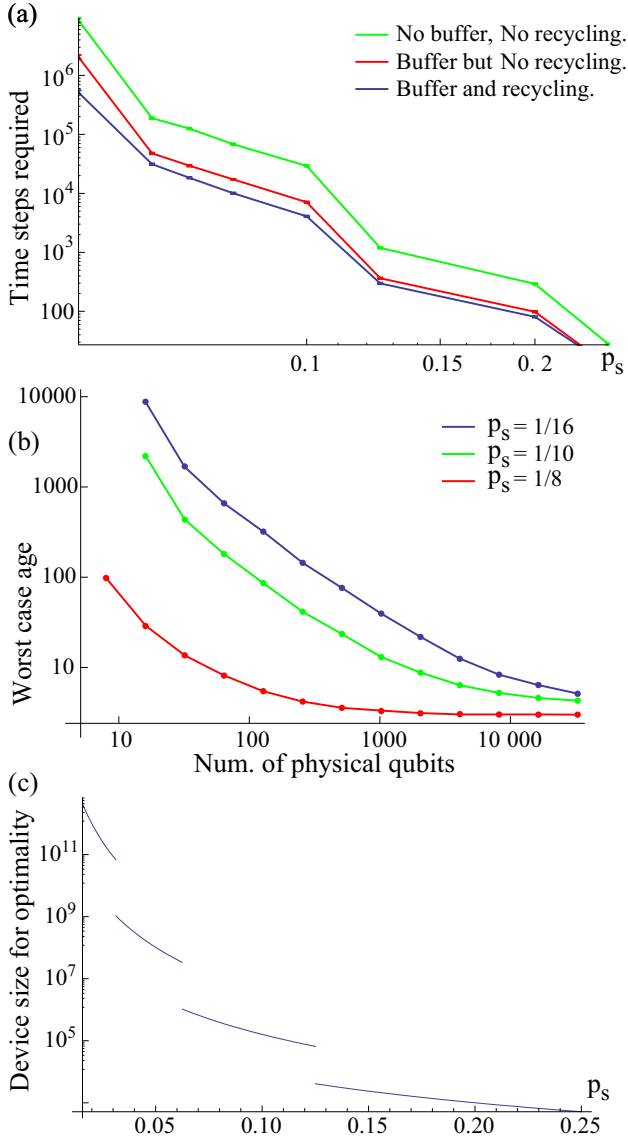}
\caption{Upper graph: The time steps required to produce a snowflake of size $1/p_S$ as a function of $p_S$. The uppermost line (green) is the case of no buffer, i.e. we insist on placing all physical qubits into the snowflake. Introducing a buffer of equal size dramatically improves the performance (middle line, red), however the introduction of recycling does not provide significant further improvement (lower line, blue). Middle graph: the `age' of the oldest entanglement between qubits within a snowflake, at the time the snowflake becomes complete, as a function of the size of total number of physical qubits. For devices consisting of at least $(\frac{2}{p})^{ \lceil \log _2 \frac{1}{p} \rceil }$ qubits, the age is simply $\lceil \log _2 \frac{1}{p} \rceil$ where $\lceil x \rceil$ is a ceiling function to denote a minimum integer which is not less than $x$. Lower graph: desired device size in order to keep errors within order $\log(1/p_S)$. The $y$-axis can be read as a cost factor to enable QIP with a probabilistic technology, as compared to a completely deterministic machine.  }
\label{fig:firstGraphs}
\end{figure}

It is efficient to grow the snowflake structure in phases. In the Phase I, we begin from product state qubits and aim to create a snowflake incorporating $1/p_S$ qubits in total. The strategy which proves the most efficient is to fuse snowflakes in pairs, each pair being matched in size. Note that this is in the spirit of the MODESTY/GREED approaches identified by Eisert {\em et al} for growth of linear graph states~\cite{Gross:2006p40}, however here we assume that the physical technology is capable of performing multiple operations in parallel (as would be the case for optical measurement-based entanglement). Snowflakes are fused at their core (see Fig~\ref{fig:simpleFusion}(c)). In the figure we depict the kind of fusion that results from a {\em parity projection}, i.e. a projector into the two-qubit subspace of given parity, since this is the type of EO that has been mostly commonly proposed~\cite{BM01a, PYF01a, PYF02a, KLM01a, BPH01a, BK01a}. If the fusion fails, then we choose to reset the complete structure back to product state qubits. This guarantees that the eventual size $1/p_S$ object will be a perfect binary tree, since it is the result of an unbroken chain of successes. It may seem wasteful to discard the relatively complex graph states that remain after a single failure. But we have found that the use of recycling, whereby one attempts to reuse such fragments, is not helpful in this Phase: we would obtain the target $1/p_S$ object only very slightly more rapidly, with the cost that there are now a random number of errors accumulated in the structure (see Fig.~\ref{fig:firstGraphs}). 

It is interesting to note that the process depicted in Fig.~\ref{fig:simpleFusion} is making aggressive use of parallelism: For $p_S\ll 1$, in a typical time step the majority of qubits must be in either the separable state or the two-qubit snowflake (the left most blocks in the Figure). Thus most of these qubits will be designated for an entanglement attempt in the next round of (simultaneous) entanglement operations.

We find that it is important to employ a buffer, i.e. the total number of qubits available in the device should be larger than the target size $1/p_S$. Otherwise, the growth process will repeatedly `get stuck' while one waits for the emergence of a snowflake to match the size of the present largest. To mitigate this effect, it suffices to have a buffer equal to the size of the desired snowflake, i.e. total number of qubits should be $\ge 2/p_S$. However, there are advantages to using far larger buffers, as noted presently.

\begin{figure}[!]
\includegraphics[width = \columnwidth]{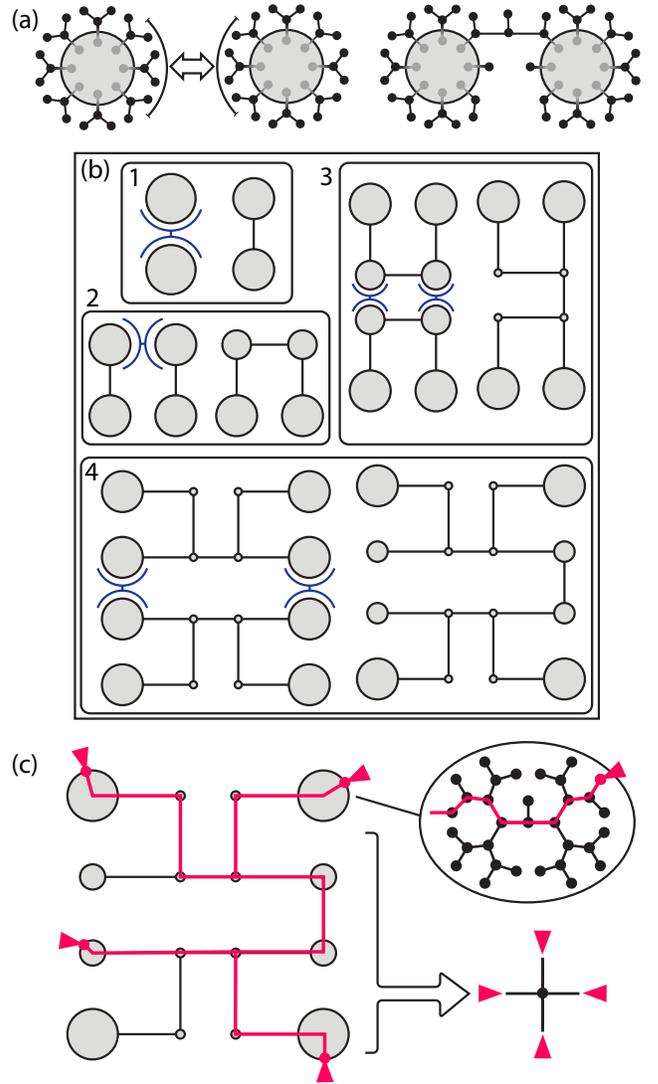}
\caption{Snowball growth in Phase II. Upper figure: A portion of the external nodes of each component object is allocated to the role of fusing to a specific partner object. On success, the new entity contains significantly more qubits. Panels $1$ to $4$ depict successive steps in growing a snowball that is large enough for a use in a subsequent cluster state synthesis (see text). In step $1$, two snowflakes of size $1/p$ are fused to obtain an object with $1.55/p$ qubits; in subsequent steps the resultant contains $2.27/p$, $3.15/p$ and $4.07/p$ qubits respectively.}
\label{fig:snowBall}
\end{figure}

Having obtained snowflakes of size $1/p_S$, we then proceed to a Phase
II where these snowflakes are combined into larger but more loosely
defined objects that we call a {\em snowballs}. A snowball is created by
attempting, in parallel, to fuse pairs of perimeter nodes of two
snowflakes (or, smaller snowballs), see Fig.~\ref{fig:snowBall}
(upper). If at least one such fusion succeeds, then the two objects are
successfully connected; thus the probability of this outcome is thus not
limited to $p_S$. Instead it is a number to be obtained by numerical
optimization. As shown in Figure~\ref{fig:snowBall}, we find
that a snowball  comprised of $4.07/p_S$ qubits can be obtained from
$16$ snowflakes of size $1/p_S$ with a probability of at least $2.31\%$ that is independent of $p_S$. A snowball of this size is a resource that enables the final phase. 

Finally, in Phase III the large snowballs are fused to form the ultimate
graph state. There are several tactics that one could employ here,
depending on the desired target state. To take a concrete example we
assume that the target is a canonical two-dimensional square cluster
state (as depicted in Fig.~\ref{fig:simpleFusion}(a)). We employ a very
basic strategy for generating such a cluster: fuse snowballs in a square
lattice, and then remove extraneous nodes. Specifically, we take
snowballs size of $4.07/p_S$ as described in the previous Phase,  and
commit a quarter of all nodes to the task of fusing to each of the four
neighboring snowballs. Then we find that the probability of achieving at
least one successful fusion between two specific adjacent snowballs is
$0.639$. Since this is significantly above the perculation threshold of $\frac{1}{2}$, it follows from the treatment in Ref.~\cite{Browne:2008p372} that the resulting imperfect cluster state will embed a perfect cluster state of somewhat smaller size (the scale factor, being independent of $p_S$, does not affect our claim of logarithmic error scaling).


To track error accumulation, we first note that graph states can be defined as states stabilized by the Pauli operators $X_i \prod_{j \in\mbox{Nbgh}(i)} Z_j$, where $\mbox{Nbgh}(i)$ denotes the neighbourhood of vertex $i$ in the graph labeling that specific graph state. A direct result of this definition is that measurements of Pauli operators made on graph states must result in other stabilizer states, which are equivalent up to local operations to smaller graph states. The relevant transformation rules were discovered independently by Hein \textit{et al}\cite{HEB04a} and by Schlingemann\cite{S01a}. Of particular consequence to graph state growth schemes are the effects of $Y$- and $Z$-basis measurements. $Y$-basis measurements complement the edges between neighbours of the measured vertex, which is removed along with any edges connected to it, while $Z$-basis measurements alter the graph simply by removing the measured vertex and any edges attached to it. $Z$ measurements can be used to remove unwanted qubits, leaving only connected paths, while $Y$ measurements can then be used to contract the path between two nodes, removing the intermediate qubits to leave a direct edge between the two nodes.

The large resource overhead required to deal with non-deterministic entangling operations can cause error accumulation to balloon, as local errors on each of the qubits used in the growth phase can be propogated to qubits in the final graph state when these ancillae are measured out. Previous schemes have relied on long paths connecting nodes in the final graph state, which scale linearly in $1/p_S$, and so the probability of avoiding error once the entire path is measured out is exponentially small in $1/p_S$. In the snowflake scheme, however, the maximum path length between nodes in the final graph state is only logarithmic in $1/p_S$. As snowballs form tree-like graphs, in order to disentangle all unwanted branches it suffices to perform a $Z$ measurement at the cutting points, which scale linearly with the path length, leading to only a polynomial decrease in success probability. Even in the worst case, the maximum path length between nodes is $10 \log_2(\frac{1}{p_S})$.

Concerning error accumulation from decoherence during snowflake growth (see Fig.~\ref{fig:simpleFusion}), one might have the following concern: since our strategy is to store a small snowflake until an equally sized partner emerges, it is possible that parts of the eventual large snowflake will be very `old' and may therefore have suffered significant decoherence while `waiting'. This would indeed be the case if a small fixed set of physical qubits were committed to the production of each snowflake. However, in reality rather than `walling off' parts of the device to produce individual snowflakes, instead all physical resources associated with snowflake growth would be shared. Thus in a full scale quantum computer, there would be numerous snowflakes of each size in existence simultaneously -- then, it will never be the case that a specific snowflake waits for a partner, and the age of the oldest entanglement relationships within a given snowflake is only a logarithmic function of the snowflake's size -- i.e. it is only ~$\log(1/p_S)$

The overall error accumulation in the ultimate graph state (e.g., cluster state) is then merely a logarithmic function of $1/p_S$, so that there is no fundamental difficulty with errors to make this approach impractical. It only remains to assess the resource scaling $1/p_S$ in order to gauge what values of $p_S$ might be tolerable in a realistic system. In the lowest graph of Fig.~\ref{fig:firstGraphs} we show the size of device needed to produce one complete snowflake of size $1/p_S$ {\em per time step}. This is the threshold where the error accumulation due to `age' of snowflakes becomes merely logarithmic in $1/p_S$, and therefore one can interpret this as {\em the scaling cost needed in order to make a probabilistic machine function similarly to a deterministic device}. As can be readily seen, the factor necessary to support $p_S<1/8$ is already high, in excess of $1000$. While such numbers might be achievable with a sufficiently dense technology, below the $p_S=1/32$ level the cost rapidly becomes unfeasible, exceeding $10^{12}$ for $p_S=1/ 64$.

In conclusion, we have introduced an error-minimising protocol for creating large entangled states using single-qubit nodes together with entangling operations (EOs) that succeed only with probability $p_S\ll 1$. This protocol makes efficient use of parallelism and bounds the error accumulation within a logarithmic function of $1/p_S$. We show how large a machine using failure-prone EOs must be, in order to compete with a machine based on deterministic EOs. 

\textit{Acknowledgements: }This research is supported by the Engineering and Physical Sciences Research Council through the QIP IRC (www.qipirc.org), and by the
National Research Foundation and Ministry of Education, Singapore. SCB acknowledges support from the Royal Society. YM is supported by the Japanese Ministry of Education, Culture, Sports, Science and 
Technology.

\bibliographystyle{apsrev}


\bibliography{bibFileVersion2}

\end{document}